# Cool Dark Cosmic Dust as a Reservoir of Absorbed Energ

Irakli Simonia

Georgian National Astrophysical Observatory
E-mail: ir_sim@yahoo.com

**Abstract.** The conception of cool dark cosmic dust has been proposed. The process of accumulation of absorbed energy by dust of such kind is considered. The conception of accumulation horizon is introduced. The possible role of cool dark cosmic dust is discussed for solution of the problem of dark matter. The other aspects of the problem are also considered.

**Introduction**

It is known that cosmic dust is under the action of electromagnetic radiation and fast particle fluxes. Stars of different spectral types and luminosity, various non-stationary objects, supernovas, pulsars, active galactic nuclei and other objects are the sources of shortwave electromagnetic radiation and of the particles of different energies interacting with cosmic dust in circumstellar, interstellar and intergalactic spaces. The interaction of radiation with dust causes a number of well-known phenomena, including absorption, scattering and luminescence. The problems of absorption and scattering of radiation by cosmic dust are studied rather well by different authors, e.g. Li (2005), Drain (2003). Photoluminescence of dust of reflection nebulae and other objects is studied by Witt (2004), Duley et al. (1997), Hendecourt et al (1986) and others.

Dust particles of different sizes, when absorbing the passing energy, transform it, in many cases, into heat or into radiation of optical range – luminescence. One or another peculiarity of the result of interaction between radiation and dust, including the formation of refractory mantle, the enrichment of dust particles with impurity ions, the change of the state of aggregation and the peculiarities of optical luminescence and IR fluorescence spectra will depend on energy of radiation, on the size and chemical composition of dust particles and on other factors. However, by the above-said it is assumed that there is an interaction of dust particles with UV radiation and with electrons of relatively small energies. In my opinion, there is also another case, namely, the interaction of cool dust with gamma rays, with hard X- ray and with different high-energy particles, including relativistic particles. I think that the atoms of cool dust matter absorbing the energetic particle, will make transition to metastable state and can stay there rather long in conditions of low temperatures and the absence of collision with other dust particles. Thus, cool cosmic dust can serve as a "reservoir" of the absorbed energy, and in any particular case, each dust particle (microreservoir) will keep strictly determined amount of absorbed energy. I advance my own hypothesis that the cool circumstellar, interstellar and intergalactic dust hides a certain amount of energy radiated by different sources. The given paper deals with the consideration of the proposed hypothesis.

## "Frozen" phosphorescence of cosmic dust

Prinsgame (1951) noted that for atomic systems being under the conditions of rather low temperatures, the transition of atoms from metastable state M to excited state F is not practically observed. At the same time, when the light is absorbed, the transitions from ground state N to the state F, as well as from F to M state are made without difficulty. This means that the phosphorescence of the corresponding matter is excited and "frozen". If then we increase the temperature of the matter, the absorbed energy is released and begins to radiate as a bright flare, without a new excitation. Thermoluminescence is the indication of "frozen" phosphorescence in certain temperature intervals. Koike, et al. (2002) and Koike, et al. (2002) irradiated the minerals of forsterite, orthoenstatite, olivine, crystalline silicone, etc. by gamma-ray and fast neutron fluxes under the conditions of liquid nitrogen temperature. The following quick heating of the irradiated minerals up to the room temperature led to their thermo luminescence in long-wavelength region of optical spectrum. Under the conditions of low temperatures, the samples of mentioned minerals of millimeter dimensions accumulated the energy of absorbed gamma photons and fast neutrons. It is obvious that at further heating of samples, the atoms from metastable state returned to excited state with further reradiation of the energy in optical spectrum. Koike, et al (2002 a, b) studied the case of frozen phosphorescence of minerals, trying to explain the nature of Extended Red Emission of reflection nebula. In my opinion, the phenomenon of frozen phosphorescence of dust has much more wide and universal significance in the process of evolution of the Universe matter. I think that the process of "freezing" (accumulation) of the energy absorbed by fine-disperse dust particles takes place permanently in cosmic medium. Various sources of hard electromagnetic radiation and fast particle fluxes; stars of different spectral types and luminosities, including non-stationary stars, pulsars, nova and supernova, active galactic nuclei, etc., irradiate permanently circumstellar, interstellar and intergalactic dust. This irradiation takes place at every chosen moment of time. The energy of absorbed particles is accumulated by the atoms of the matter of dust particles making transition to the metastable state as a result of absorption process. Each dust particle of microscopic size with atoms of its matter in the metastable state, is a sort of microreservoir of absorbed energy. Under the conditions of stable low temperatures $T < 77$ K, the energy accumulated by dust particles can retain for million years and only as a result of quick heating of dust or collision, the absorbed and "frozen" energy will be released as an optical radiation – thermoluminescence. It is natural that well-heated dust will not be able to accumulate the absorbed energy, while the dust being at significant distance from IR radiation sources will accumulate and keep the absorbed energy without difficulty under the conditions of stably low temperatures. Taking into account the motion of dust, the possibility of its heating by UV radiation, the frequent collision processes of dust-dust or dust-gas type, and other dynamical processes, one can conclude that simultaneously with the process of absorption and accumulation of energy by dust, the process of the release of accumulated energy by dust particles will take place. Thus, in cosmic medium two opposite processes will take place permanently: the process of accumulation of energy by dust and the process of the release of energy by dust. If these processes are balanced, it can be expressed as:

$$nE_{abs.} = n_1(E_{rad} + E_{int}) \qquad (1)$$

where $n = n_1$ is the number of dust particles in some separate closed volume or in the Universe as a whole, $E_{abs.}$ is the absorbed energy, $E_{rad}$ is the radiated energy in the form of thermoluminescence and $E_{int.}$ is the energy consumed for the internal processes.

At the same time, in the Universe at each chosen moment of time t, there will be a finite number of dust particles $n_2$ neither accumulating nor releasing the energy. This dust has already absorbed a certain amount of energy and keeps it until a certain moment of time $t_1$ – the moment of external effects (heating, collision, etc.). The consideration of numerical quantity of such dust and of the corresponding accumulated energy is difficult even if we know the exact value of gas/dust ratio. The dust of such kind, that keeps the absorbed energy, can be named dark dust.

The cosmic dust can be divided into two classes: a) the dust accumulating the energy, b) the dust does not accumulating the energy. Such division of dust is caused by a number of factors, including chemical-mineralogical composition of dust particles (the atoms of some organic compositions have quite deep metastable levels), dust temperature, size of dust particles, density of dust clouds, etc. However, I think that the factor of temperature is more important. Probably, in the Universe there is a certain boundary or temperature limit, lower of which the matter, including the dust, becomes able to accumulate the absorbed energy. Such limit, probably, is near T= 77 K. I propose to call this limit the horizon of accumulation.

The dust having the temperature T< 77 K, i.e. the dust lying below the horizon of accumulation, can accumulate the absorbed energy. The dust having the temperature T> 77 K, i.e. the dust lying above the horizon of accumulation, cannot accumulate the absorbed energy. Obviously, the dust particles of micron size or large sizes can accumulate the absorbed energy.

The dark dust is below the horizon of accumulation. One cannot exclude also that the interstellar gas in the form of the complex of collisionless polyatomic molecules (e.g. PAHs), having the temperature $T < 77$ K, i.e. lying below the horizon of accumulation, will as well keep the absorbed energy for a long period of time. Thus, below the horizon of accumulation: 1. The cosmic dust (possibly the gas too) can keep the absorbed energy during a rather long period of time. 2. The energy accumulated by dust is still impossible to be registered. 3. The energy released as a thermoluminescence (the result of dust heating) will be also difficult to be registered due to the relatively small quantum yield. So, the cool dark dust can hide a certain part of energy of the Universe, not giving us, in fact, any possibility to detect this hidden energy even at the moment of its release. Registration of thermoluminescence of scattered dust, excited by gamma-rays, by the fluxes of fast neutrons and other particles, is a difficult instrumental task.

On the other hand, probably there exists a counterargument - extended Red Emission of reflection nebulae and other objects. The energy absorbed and accumulated by cool dust can play a definite role in solution of the problem of dark matter. For a ground-based observer there are two limitations: a) the accumulated energy is hidden by cool dust, b) the released energy is difficult to be registered.

The processes of absorption and accumulation of energy by cool dust with subsequent reradiation of energy kept for some period of time, take place permanently in the Universe. At the same time, the above-mentioned limitations are also permanent.

If in the intergalactic space there is dust of micron size under the conditions of super-low temperature ($T < 8$ K), probably it can keep the accumulated energy during extremely long periods of time. Such dust can be a carrier of a definite part of non-registered energy. The registration of the dust of such kind is problematic at present.

I think that cool intergalactic dust particle being below the horizon of accumulation and keeping the energy of at least one gamma-photon is an important element in the process of evolution of the Universe matter, requiring a special attention not only from the point of view of astrophysics, but from the viewpoint of cosmology. Let us list now some objects, the radiation of which can be accumulated by cool dust. Among these objects are: supernovas, pulsar PSR 0531+21 in Crab Nebula, pulsar in constellation Vela, dense cloud in constellation Ophiuchus, quasar 3C 273, etc. Cool dust accumulates the energy of absorbed fast neutrons and other particles, as well as soft and hard gamma-rays from the sources of the mentioned type. The dust being below the horizon of accumulation will also interact with galactic diffuse gamma-rays, accumulating the absorbed energy and keeping it until the moment of quick heating. Only the dust being at small distances from the corresponding sources, will accumulate the energy of absorbed neutrons. At the same time, at any point of interstellar or intergalactic medium, the cool dust will keep the absorbed energy of other particles or gamma-rays during the long periods of time.

I propose to express the total energy accumulated by dust as

$$E = KNF \qquad (2)$$

where K is the coefficient expressing the probability of the process of accumulation of the energy absorbed by dust, $K = f(T,D)$, where T and D are the temperature and diameter of a dust particle, respectively. In general case, $0.9 > K > 0.1$. When $K = 0.1$, it means that only 10% of dust is at the temperature $T < 77$ K, i.e. below the horizon of accumulation. N is the total number of dust particles in the corresponding space region. F is the power of radiation absorbed by dust particles.

For the case: radiation source – interstellar dust - the ground-based observer, it is appropriate to consider the total number of dust particles in a certain column with 1 cm$^2$ cross-section, located on the line of sight. The proposed formula (2) will be valid for spatially fine or rarefied dust structures. It is natural that, the electromagnetic or corpuscular radiations of corresponding energies can penetrate into the inner layers of dense dust clouds only to some extent.

Let the conditional source of gamma rays be at the distance of 2 kpc from the ground-based observer. Let, as well, $10^9$ dust particles per 1 kpc be concentrated in the column with 1 cm$^2$ cross-section located on the line of sight. Let us take $K = 0.1$. Gamma-photons from the source have 2.23 MeV energy. The energy of absorbed gamma photons is accumulated by 10% dust being on the line of sight. According to formula (2), the total energy accumulated by dust will be $E = 8.10^2$ erg/sec. Calculation made only for one conditional source and for photons of chosen energy shows that cool dust is able to accumulate a considerable energy. If we make calculation for several real sources of hard radiation located at different distances, the energy accumulated by cool dust can be remarkable.

Speaking about the possibility of indirect detection of cool dust – microreservoirs of absorbed energy, one should remember that the dust of such kind can be responsible for interstellar absorption at corresponding wavelength in shortwave region of spectrum. Cool dust particles of micron and large size interacting with gamma rays, hard X- ray, neutrons and other particles of high energy accumulate a significant energy and keep it for a long period of time.

**Conclusion**

Cool interstellar and intergalactic dust accumulating the absorbed energy of hard electromagnetic and corpuscular radiation, plays an important role in evolution of the Universe. The above-considered processes clearly show the variety of possible interactions of the matter with radiation, and one form of matter with its another form. Cool dust, lying below the horizon of accumulation is the keeper of absorbed energy for a long period of time. The "frozen" phosphorescence of cool (dark) dust particles can serve as a key for solution of a number of astrophysical and cosmological problems. I offer to have discussions on the proposed hypothesis.